\documentclass{iopart}
\usepackage{graphicx}
\usepackage{float}
\begin{document}

\title[Measurements of a dynamical temperature in turbulence.]{Fluctuation-dissipation relation on a Melde string in a turbulent flow, considerations on a ``dynamical temperature''.}
\author[GGN]{V Grenard, N B Garnier and A Naert.}
\address{Universit\'e de Lyon, Laboratoire de Physique, \'Ecole Normale Sup\'erieure de Lyon,\\
46 All\'ee d'Italie, 69364 Lyon Cedex 07, France.}
\ead{Antoine.Naert@ens-lyon.fr}
%{Nonequilibrium and irreversible thermodynamics}
%{Fluctuation phenomena, random processes, noise, and Brownian motion} 
%{Statistical mechanics of classical fluids}
%\maketitle
\begin{abstract} 
We report on measurements of the transverse fluctuations of a string in a turbulent air jet flow. Harmonic modes are excited by the fluctuating drag force, at different wave-numbers. This simple mechanical probe makes it possible to measure excitations of the flow at specific scales, averaged over space and time: it is a scale-resolved, global measurement. We also measure the dissipation associated to the string motion, and we consider the ratio of the fluctuations over dissipation (FDR). In an exploratory approach, we investigate the concept of {\it effective temperature} defined through the FDR. We compare our observations with other definitions of temperature in turbulence. From the theory of Kolmogorov ($1941$), we derive the exponent $-11/3$ expected for the spectrum of the fluctuations. This simple model and our experimental results are in good agreement, over the range of wave-numbers, and Reynolds number accessible ($74000 \leq Re \leq 170000$). 
\end{abstract}
\section{Introduction}
Turbulent flows exhibit a notoriously complex and unpredictable dynamics: they present a huge number of degrees of freedom, and their dynamics are both far from equilibrium and dissipative \cite{landau6,monin_yaglom,frisch}. The kinetic energy injected at large scale by shear instability mecanisms is dissipated into heat by the molecular viscosity at small scales. That is, dissipation and injection scales are distinct. Therefore, a transport process through scales is necessary for a flow to be stationary. It is suspected that instability mechanisms associated with non-linearities generate harmonics, therefore transfering energy to smaller scales almost without dissipation. An equivalent picture would consist in vortices stretching each other in such a way that a non-zero energy transfer occurs toward smaller scales. This picture of \emph{cascade} process was first proposed by Richardson \cite{richardson22}. The cascade stops approximately in the range of scales where the viscosity becomes efficient to damp velocity gradients. In the late thirties, Kolmogorov derived from this idea a phenomenological theory accounting for the fluctuations of various observables in fully developed turbulence \cite{k41}. In the present work, we are neither concerned by the large (energy injection) scales, nor by the small (dissipation) scales, but by the intermediate range. In this intermediate inertial range, we study the transport process through scales, expected to be universal. Instead of scale $l$, one often refers to the wave-number $k=2\pi/l$.\\
The control parameter of the flow is the Reynolds number: $Re=\frac{V L}{\nu}$, where $L$ is the macroscopic scale of the flow (integral scale, or correlation length), $V$ is a characteristic shear velocity at large scale, and $\nu$ is the kinematic viscosity of the fluid. It is also the mean ratio of the inertial by the dissipative contribution of the forcing over a fluid particle. Interesting predictions were derived by Kolmogorov ($1941$), that we use in the following. Especially, the range of scales over which fluctuations occur scales as $Re^{3/4}$. The prediction for the exponent of the power spectral density as $\langle|\tilde{v}|^2\rangle \; \propto \; k^{-5/3}$ is among the most famous successes of this theory \cite{landau6,monin_yaglom,frisch}.\\
Our experimental system is discribed in detail in the next section. It is a thin string held by its ends at constant tension across a turbulent flow. To formalize briefly, it is an oscillator with multiple resonances, coupled to a particular 'thermostat': the turbulent flow. This string is used to probe the inertial range of a flow of high enough Reynolds numbers. The device is 'calibrated' by measuring the average (complex) response to an external perturbation, and then used to measure the free fluctuations caused by turbulence alone. Measurement of the displacement $r(t)$ caused by the turbulent forcing $f(t)$ is performed with small piezoelectric transducers. We measure the average response, i.e. the displacement on one end caused by a known broad band forcing on the other end. Then, measurements of the displacement on one end alone give information on the forcing fluctuations. Our study goes a step forward, in an exploratory way. Knowing the average response function of the string and measuring $r(t)$, we invoque a version of the Fluctuation-Dissipation Theorem extended out of equilibrium, to define an effective temperature of the turbulent flow. This effective temperature happends to be scale-dependant.\\
In this work, fully developped turbulence is addressed from the point of view of statistical mechanics. We first recall one important break-through: the statement of the Fluctuation-Dissipation Theorem (FDT). Consider a pair of conjugate variables (displacement $r$ and force $f$) of a small system in thermal contact with a large heat reservoir. In the present case the small system is the string, coupled to the turbulent flow which is the reservoir. Displacement $r$ and force $f$ are conjugate in the sense that their product is the work exerted by the flow on the string. The theorem originates from the idea that spontaneous fluctuations $r(t)$ should have the same statistical properties as the relaxation of $r(t)$ after the removal of an external forcing perturbation. The main hypothesis needed to derive this theorem are: --~linear response between $f$ and $r$, --~thermal equilibrium between the system under consideration and the thermostat, --~thermal equilibrium of the thermostat itself. The response function $H_{r,f}$ is such that: $r(t)=\int_{-\infty}^{t}H_{x,f}(t-t')f(t')dt'$. Equivalently it can be written in the Fourier space as: $\tilde r(\omega)=\tilde H_{r,f} \; \tilde f(\omega)$. Under some hypothesis, the fluctuations of $r$ (its 2-times correlation function) are linked by a very simple relation with the dissipative response of the system to a perturbation of the conjugate variable $f$ (imaginary part of the average response function). It is simply proportional, and the coefficient is nothing but the temperature multiplied by the Boltzman constant: $k_{\rm B} T$ \cite{kubo}. The validity of the hypothesis has to be discussed in each case. If they are satisfied, the correlation function of the spontaneous fluctuations is proportional to the response function, i.e. the factor is unique and constant. Moreover, this factor is the same for all couples of conjugate variables, and this factor is $k_{\rm B} T$, where $T$ is the temperature of the system. The Boltzman constant $k_{\rm B} \simeq 1.38\;10^{-23}JK^{-1}$ is an universal constant. This relation can be expressed in spectral variables: 
\begin{equation}
  \label{fdt}
  \langle|\tilde r(\omega)|^{2}\rangle=\frac{2\;k_{\rm B}T}{\omega}\;Im[\tilde H_{r,f}(\omega)]. 
\end{equation}
In this expression of the FDT, $\langle|\tilde r(\omega)|^{2}\rangle$ is the power spectral density of the fluctuations of the displacement $r$, as $\tilde H_{\rm r,\rm f}(\omega)$ is the response function on $r$ to the conjugate variable $f$. Because the string is very thin, the drag is purely viscous. It is therefore proportional to the velocity, which is in quadrature with the displacement. The dissipation is therefore proportional to the imaginary part of the average response function: $Im[\tilde H]$.\\
In the perspective of constructing a non-equilibrium thermodynamics, the FDT has been reconsidered by L.~Cugliandolo and J.~Kurchan, while investigating amorphous materials relaxing after a thermal quench through the glass transition \cite{cugliandolo93,cugliandolo97}.\\ 
We present in the following an exploratory approach of the question of turbulent fluctuations using their extended formalism. The Fluctuation-Dissipation Ratio (FDR) can be rewritten: 
\begin{equation}
  \label{fdr}
  \frac{\omega\;\langle\tilde r(\omega)^{2}\rangle}{Im[\tilde H_{\rm r,\rm f}(\omega)]}=2\;k_{\rm B} T_{\rm eff.}(\omega),
\end{equation} 
where the temperature is replaced by an 'effective' temperature $T_{\rm eff.}$, function of frequency $\omega$. The frequency dependence of $T_{\rm eff.}$ expresses the fact that different degrees of freedom are not at equilibrium with each other, resulting in internal energy fluxes.\\
In other words, in our system, each (independent) mode of the string couples to (non-independent) scale of the flow. As the flow is stationary, we average our measurements on time, and finally obtain the frequency dependance of $T_{\rm eff.}$ as defined by equation~\ref{fdr}. Measurements of the fluctuations of the string give Fourier components of the excitation of the flow. We measure independently the fluctuations, and the complex average response function to a specified excitation, in a way discussed below. We propose to analyse these measurements with the criteria discussed above.\\
The paper is organised as follows. The next section describes the experimental setup, turbulent flow properties, and the setting of the string. General properties of a vibrating Melde string are also discussed. The measurements are shown in section \ref{sec:3}: response, fluctuations, and the Fluctuation Dissipation Ratio of this system. In section \ref{sec:4}, we derive from Kolmogorov's theory a simple scaling model for the fluctuations of the drag, and therefore the FDR, which accounts for the exponent observed in the whole range of accessible $Re$. The section \ref{sec:5} is devoted to a discussion of our results, especially in comparison to several definitions of temperature in turbulence proposed in the literature. 
\section{The Melde string and the experimental setup} \label{sec:2}
\begin{figure}[h]
  \begin{center}
    \includegraphics[scale=.55]{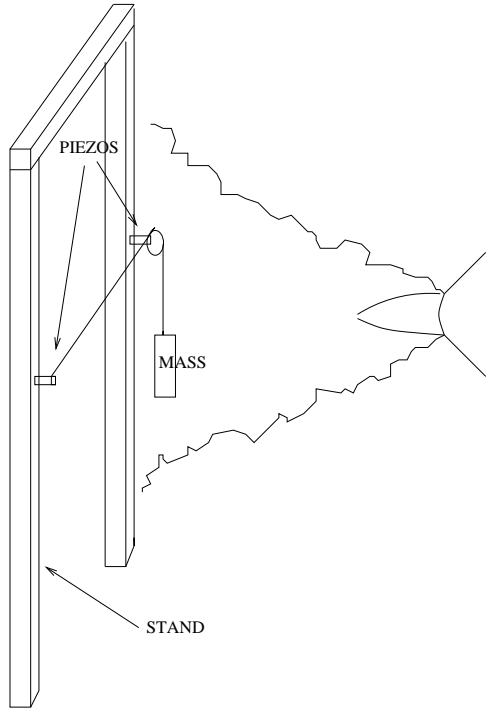}
  \end{center}
  \caption{Eperimental setup: the thin steel wire is pulled across a turbulent air jet by a $4$ Kg weight on a rigid stand. Piezoelectric transducers are in mecanical contact with the wire at each end.}
  \label{f.1}
\end{figure}

The experimental setup is sketched in Fig. \ref{f.1}. A turbulent air jet originates from a nozzle of diameter $5$ cm. The flow facility we used is thoroughly described in \cite{marcq01}. A thin stainless steel string of length $60\,$cm is located $2\;$m downstream the nozzle, perpendicular to the axis of the flow. At this distance, the length of the string is about the diameter of the turbulent jet. The displacement of the string is measured using piezoelectric multi-layer ceramics at each end of the string. A piezo is deformed by a voltage. Reciprocally, if the ceramic in compressed, a voltage is generated. The relation between voltage and deformation is linear, and the frequency response is almost flat in the frequency range we consider here. It can be used as actuator or sensor. We have two piezos, one on each end of the string. The two different measurements we perform are the following. 1)~complex response function: one (input) piezo is feeded with a white noise voltage through a power amplifier. The source is that of a HP$3562$A signal analyser. Standing transverse waves appear in the string, weakly perturbed by the turbulent fluctuations. Mecanical displacement on the other end is transformed into a voltage by the other (output) piezo. It must be amplified, and both input and output voltages are recorded synchronously with a $24$ bits A/D converter. The acquisition frequency is $50$ kHz. We call response the time averaged ratio of the voltage amplitudes on input and output piezos, recorded simultaneously. Voltages {\it in} and {\it out} are proportional respectively to the displacement and the constraint (on the piezos). The dimension of the actual response is the inverse of a stiffness, as what we measure is the ratio of voltages. Dimentional prefactors are omited for simplicity, as they are constant for the same setup (string and transducers). The diameter of the string is $100\;\mu$m, less than the viscous scale of the flow which is about $\eta \simeq 170\;\mu$m at the largest $Re$ accessible.
\noindent
The equation of motion of the undamped and unforced string is a linear wave equation. Its solutions with fixed ends are standing waves $r(x,t)=A\,\cos(\omega_n\,t-k_n x)$, where $A$ is the amplitude, $t$ is time and $x$ is position along the wire. The discrete wave numbers are $k_n=n\frac{\pi}{L}$, where $L$ is the length of the string and $n$ is a positive integer. In a first approximation, the waves are not dispersive: $\omega_n=c\,k_n$, where $c$ is the phase velocity. $T$ is the tension of the string and $\mu$ its mass per unit length, $c=\sqrt{T/\mu} \simeq 300\;$m/s. With a $4\;$kg weight on one end, the string's fundamental frequency is $f_0=344\;$Hz.\\
Dissipation is mainly due to friction on air, and causes little dispersion. More precise treatment would require terms of dissipation in the wire itself and in the piezoelectric transducers that fix the ends. We neglect this, as the amplitude remains small (a few tens of micrometers) if compared to the length of the ceramic pile ($3$\,mm), or even the wire diameter ($100\,\mu m$). 
The possible coupling with compression wave is not relevant, as the range of frequency is distinct. (Compression wave speed in steel is a few thousands of m/s, larger than what we consider here: $c \simeq 300\;$m/s.) When this wire is immersed into the turbulent flow, the resonant modes are excited by the drag forcing. The quantities measured are averaged along the wire. They are therefore global in space but local in scale, or more precisely in Fourier-space. The vortices at scale $l$ are expected to excite modes of wave-number $k=2\pi/l$. In that sense, the string is acting like a mechanical spectrometer, almost exactly like a Fabry-Perot interferometer.
\section{Measurements}\label{sec:3}
\begin{figure}[h]
  \begin{center}\includegraphics[scale=.45]{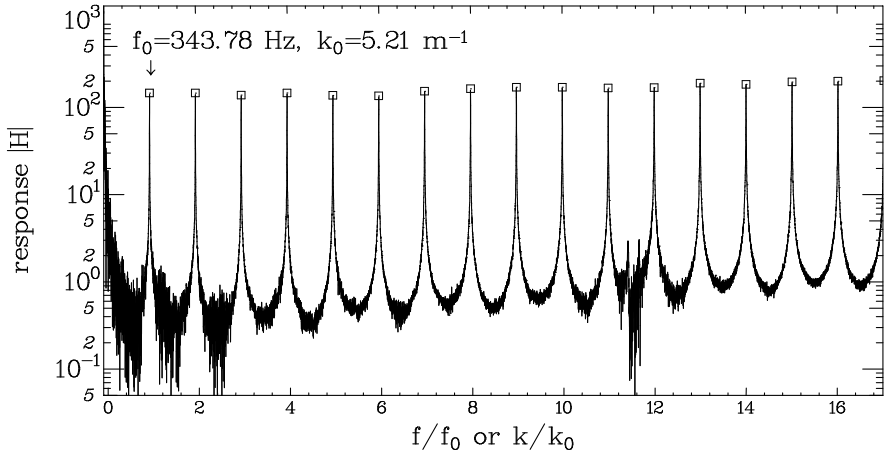}
  \end{center}
  \caption{Modulus of the response function versus the harmonic number, at $Re=154000$. The abscissa is given in non-dimensional coordinates, normalised by the fundamental frequency.} 
  \label{f.2}
\end{figure}

\begin{figure}[h]
  \begin{center}
    \includegraphics[scale=1]{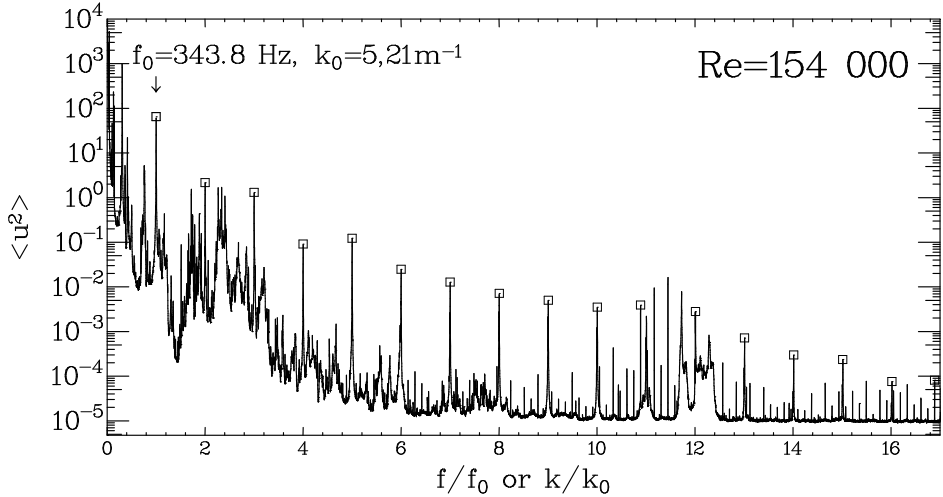}
  \end{center}
  \caption{Spectrum of the resonance modes of the string excited by turbulent drag fluctuations, at $Re=154000$.} 
  \label{f.3}
\end{figure}

\begin{figure}[h]
  \begin{center}
    \includegraphics[scale=.45]{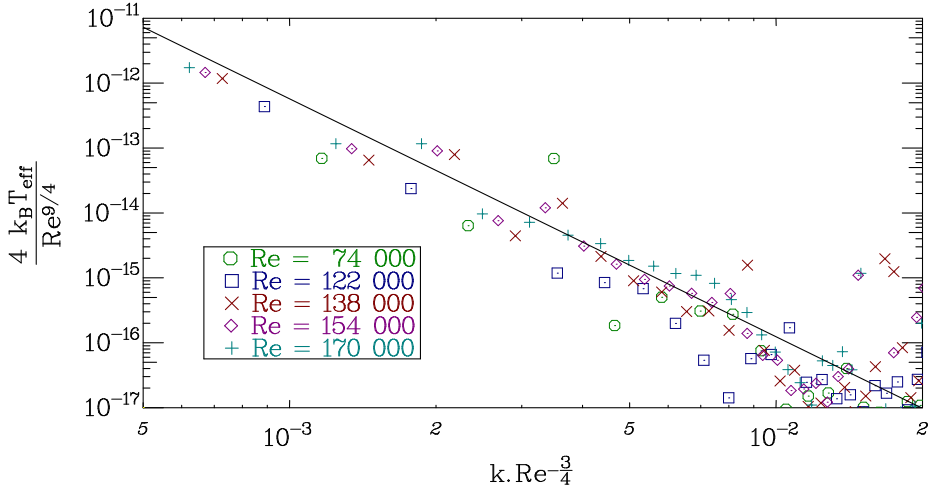}
  \end{center}
  \caption{Spectrum of the FDR, labelled as \emph{thermal agitation} per degree of freedom. Axis are rescaled with proper Reynolds number dependence, between 74000 and 170000. The size of the symbols represents the uncertainty in the determination of the maxima of the peaks. The solid line is a $\rm{k}^{-11/3}$ power-law given as an eye guide.} 
  \label{f.4}
\end{figure}

Modulus of the response function is plotted in Fig. \ref{f.2}. It shows that the resonance peaks are indeed very narrow, ensuring a very precise selection of wave-numbers: the quality factor is approximately $Q \simeq 4000$. The imaginary part of the response function is giving the dissipation.   
\noindent
The width of the peaks in the modulus is also linked to the dissipation, as well as the damping time after a perturbation. We used in the following the measurement of the imaginary part of the response, but checked that these different methods coincide. Only the resonant frequencies are considered in this study, as they are much more sensitive to the velocity fluctuations. This is especially important at large $k$, as the kinetic energy of the flow is small. Spectrum of the fluctuation excited by the turbulent drag is shown in Fig. \ref{f.3}. Fluctuations resonance peaks are clearly identified. Spurious vibrations are visible, mainly caused by the vibrations of the stand. Because the peaks are very thin, long acquisitions are necessary, as well as large windows for the FFT calculations ($150000$ points), in order to achieve a sufficient resolution ($0.33\,$Hz). The protocol we used to find the resonance frequencies, the value of the amplitude of fluctuations, and imaginary part of the response, is the following. Resonance frequency is obtained by spline smoothing each peak around the maximum amplitude of the response. Then, imaginary part is measured after being also smoothed. The amplitude of the fluctuations peaks are collected on the spectrum, after local smoothing around the maxima. 
\noindent
One can see the FDR in Fig. \ref{f.4}, called $k_{\rm B} T_{\rm eff.}$, for several values of $Re$. Uncertainties on this ratio have multiple origins. Errors indicated by the size of the symbols are those coming from the determination of the resonance frequencies. Spurious vibrations of the stand are difficult to handle: we perform measurements of response and fluctuations in the same conditions, to reduce its influence on the ratio. We believe the scattering of the points in Fig. \ref{f.4} comes mainly from the weakening of signal/noise ratio for large frequencies, simply because there is less energy in the flow at large $k$, especially at small $Re$. The only possible escape on this point is to improve the coupling between the string and the sensors.
\noindent
The wave-number has been rescaled with the internal viscous scale $\eta \propto Re^{-3/4}$. The ordinates have been rescaled by an estimated number of degrees of freedom: $(L/\eta)^3 \propto Re^{9/4}$. These $Re$ scalings are both usual consequences from Kolmogorov's theory. In other words, the \emph{``thermal energy''} $k_{\rm B} T_{\rm eff.}$ that the FDR is representing in the framework of Cugliandolo \etal's theory, is given per degree of freedom. Assuming the number of degrees of freedom is the total number of particles of size $\eta$ in the total volume is usual, but crude. A more realistic description should involve correlations between them, reducing this number. However, all the curves collapse to a single power-law with this scaling. The exponent is discussed in the following section.\\
Please note that the equipartition of energy at equilibrium would require this spectrum to be constant. There is no equilibrium between the Fourier modes, because of the energy flux through scales. Moreover, they are not independent, and probably not Gaussian. There is no reason to expect equipartition. Considering a kinematik temperature as poportional to the kinetic energy, like in the kinetic theory of gases, it would be: $T\;\propto\;\langle\tilde{v}^2\rangle$. And, because of Kolmogorov's theory it would scale as $k^{-5/3}$. The dependance we observe with our definition is much steeper. 
\section{Scaling law}\label{sec:4}
Because the susceptibility of the string is very high at resonance, the half-wave-length modes $n\lambda/2$ match with velocity structures of scale $l$ ($n$ is an integer). Therefore, the wave number of the standing wave in the string $k=n\;2\pi/\lambda$ is the same as $k=2\pi/l$. The necessary condition for this matching is resonance. It also ensures that velocities of the string and fluid equalise, which is crucial for the following argument.\\
Displacement is proportional to the drag forcing, itself proportional to velocity, as drag is viscous: the string diameter-based Reynolds number is small (about $10$). The Melde string is not dispersive: $\omega = 2\pi f=ck$, $c$ being the wave velocity. Therefore, the displacement is $r=v/\omega=v/(ck)$, and its power spectrum is: $\langle\tilde{r}(\omega)^2\rangle\;=\;\langle\tilde{v}(\omega)^2\rangle(ck)^{-2}\;\propto\;k^{-11/3}$. Because the viscous dissipation at each resonance is proportional to frequency, the FDR of Eq. \ref{fdr} is simply proportional to $c\,k\,\langle\tilde{r}(\omega)^2\rangle\;\propto\;k^{-11/3}$. Following Eq. \ref{fdr}, an effective ``thermal agitation'' defined by the FDR would be: $k_{\rm B}T_{\rm eff.}\propto k^{-11/3}$, in the inertial range of fully developed turbulence. This exponent is compatible with the spectrum we measured, as can be seen in Fig. \ref{f.4}. 
\section{Discussion}\label{sec:5}
Theoretical characterisation of turbulence in terms of temperature were proposed in the past by several authors. The temperatures as defined by T.~M.~Brown \cite{brown82} and B.~Castaing \cite{castaing96} do not depend on $k$ throughout the inertial range. The qualitative idea is that the cascade transport process is efficient enough to equalise a quantity they call temperature. In another model invoking an extremum principle, B.~Castaing proposed a definition of temperature, which might depend on scale \cite{castaing89}. In any case, none of these theories invoke the FDR. %{\bf Pope et Ching ?} 
On different basis, R.~Robert and J.~Sommeria proposed a definition of temperature \cite{sommeria91}, only valid for 2D turbulence. It is not expected to apply in a $3$D flow.\\
Now, let's consider our experimental results from the perspective of the three points of reflexion we proposed in the first section, in relation with the FDT. 1- Linear response: as we mentioned, the coupling between the string and the flow is purely viscous. Therefore, drag force is proportional to velocity: $f(t)=\gamma\,v(t)$, $\gamma$ being a friction coefficient. It is also the time-derivative of the position $f(t)=\gamma\,\omega\,r(t)$. Response is linear in $r$, but the coefficient depends on frequency. 2- Are fluctuations and dissipation proportional ? As we have seen, the measurements of the FDR are consistent with a $k^{-11/3}$ scaling, it is definitely not constant with respect to $k$. As our system is out of equilibrium but stationary, there is no time evolution like the relaxation of glasses. 3- Setting a string in a turbulent flow allows to perform measurements on a couple of conjugate force-displacement variables. We have no other set of observables to compare with, for now. \\
We may ask whether what we measure is actually a temperature, in a dynamical sense. If one assumes that each mode of the string is a harmonic oscillator, and that a harmonic oscillator at equilibrium with a bath gives the temperature of this bath through the FDR, then equilibrium between modes of the string and modes of the flow means the temperature is equal: measurements give the temperature of the flow at this corresponding scale. Such interpretation still rely on the assumption that FDR on the oscillator gives the temperature of the oscilaror: this is our working hypothesis. By equilibrium between modes of the string and the flow, we mean a 'no-flux' condition on energy. This is ensured by the high susceptibility of the string at resonance. In other words, the probe and the reservoir are in equilibrium with each other for each $k$, but equilibium is obviously not expected between one scale and another.\\
We have performed measurements on a turbulent flow, coupling to it a set of harmonic oscillators: a Melde string. At equilibrium with the flow, in the sense that each mode of the string couples with the fluid at scale $l=\pi c/\omega$. It gives informations much like a spectrometer, even though the flow itself is strongly out of equilibrium. This is true, of course, as long as the response of the string is fast enough compared to the frequencies of the velocity fluctuations. The displacement spectra are recorded at different values of $Re$, as well as the complex response of the string over an excitation (contributions of all the standing waves). \\
The matching of the string's modes and hydrodynamic structures, what we call equilibrium between the string and the flow, is still a questionable working hypothesis. However, drawing inspiration from Cugliandolo \etal's theory of non-equilibrium temperature based on the FDR, we measured the Fluctuation over Dissipation Ratio of our string in a turbulent flow, for different values of $Re$. The FDR, multiplied by an appropriate power of the Reynolds number exhibits a unique power law, when Reynolds number is between $74000$ and $170000$. The  exponent is consistent with a value $-11/3$ given by a very simple model derived from Kolmogorov $1941$ theory.
\ack
We acknowledge B.~Castaing, E.~Leveque, P.~Borgnat, F.~Delduc, S.~Ciliberto, E.~Bertin, and K.~Gawedzki for many discussions. We also thank V.~Bergeron, T.~Divoux, and V.~Vidal for corrections on the manuscript and for many discussions. Thanks to F.~Dumas for his help in the construction of positioning devices. As this system became a teaching experiment, several students contributed to this study as part of their graduate lab-course. They are gratefully acknowledged: A.~Louvet, G.~Bordes, I.~Dossmann, J.~Perret, C.~Cohen, and M.~Mathieu. We also thank the guitar maker D.~Teyssot, from Lyon, who gently gave us his thinnest E strings. 
\vspace{1cm}

\bibliographystyle{unsrt}
\bibliography{biblio3}
\end{document}